# Wavelets, Curvelets and Multiresolution Analysis Techniques in Fast Z Pinch Research

Bedros Afeyan,[a] Kirk Won,[a] Jean Luc Starck,[b] and Michael Cuneo[c]

[a] Polymath Research Inc., Pleasanton, CA
[b] Centre d'Etude de Saclay, CEA, France
[c] Sandia National Laboratories, Albuquerque, NM

**ABSTRACT**

Z pinches produce an X ray rich plasma environment where backlighting imaging of imploding targets can be quite challenging to analyze. What is required is a detailed understanding of the implosion dynamics by studying snapshot images of its in flight deformations away from a spherical shell. We have used wavelets, curvelets and multiresolution analysis techniques to address some of these difficulties and to establish the Shell Thickness Averaged Radius (STAR) of maximum density, $r^*(N, \theta)$, where N is the percentage of the shell thickness over which we average. The non-uniformities of $r^*(N, \theta)$ are quantified by a Legendre polynomial decomposition in angle, $\theta$, and the identification of its largest coefficients. Undecimated wavelet decompositions outperform decimated ones in denoising and both are surpassed by the curvelet transform. In each case, hard thresholding based on noise modeling is used.

## 1. INTRODUCTION

High yield targets for inertial confinement fusion (ICF) have been designed recently which rely on a double Z pinch configuration.[1] These novel approaches to indirect drive inertial confinement fusion energy production promise to attain fusion by next generation Z pinch machines such as those proposed by Sandia National Laboratories.[2] The idea is to attain the uniform implosion of an annular shell containing fusion fuel via intense X ray illumination in an intermediate chamber sitting between two wire array Z pinches which create the necessary mega-Joules of X rays.[1-3] To diagnose the imploding shell at the center of the middle chamber of a double Z pinch hohlraum (DZPH), an X ray backlighter is used. The X rays must enter the middle chamber from its side, and image it through a slit on the other side of the chamber at an external film plane.[4] This configuration is shown schematically in Fig. 1.

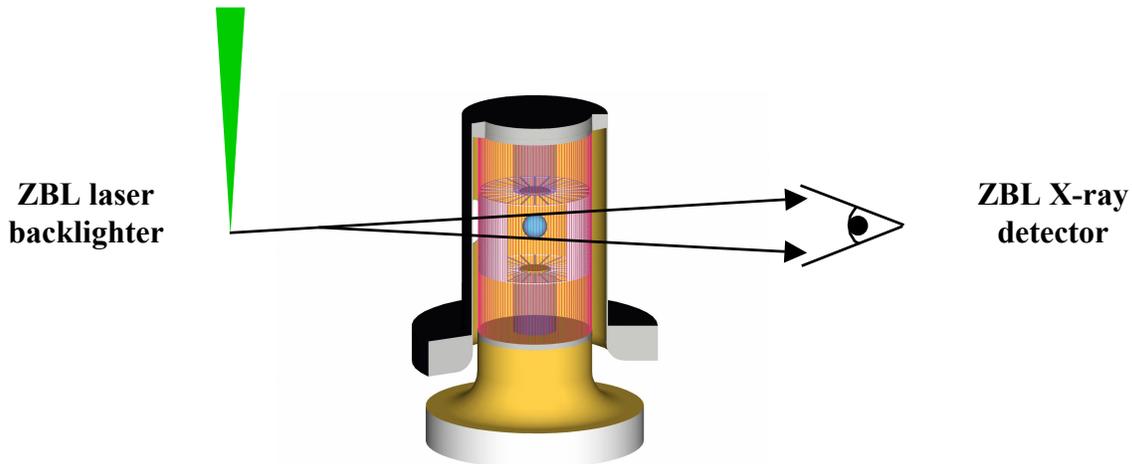

**Figure 1.** Schematic layout of a Sandia Double Z Pinch Hohlraum (DZPH) experiment showing the Z beamlet laser hitting a target, generating X rays which are then used as the backlighter[4] to image the imploding shell at the center of the middle chamber of the device.

The images are collected on a photographic plate facing the target from the opposite side. The shell implodes because of the radiation driven ablation of the shell driven by copious amounts of X rays that enter the central chamber having been created by the two imploding Z pinch wire arrays on either side, which generate mega-Joules of X rays each.[1-3] These X ray fluxes, if synchronized enough to allow for a uniform radiation distribution inside the central hohlraum chamber, will cause the uniform implosion of the fusion capsule in the center. The quantitative assessment of the degree to which this is so is our goal when we denoise such X ray backlighting generated images. The three images we will denoise and analyze are given in Fig. 2.

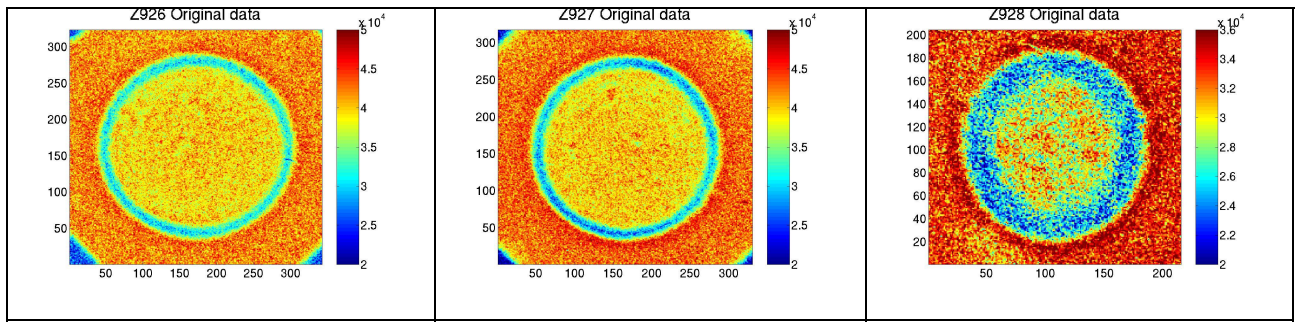

**Figure 2.** X Ray backlighting generated images of double Z pinch compressed hollow shells in Z shots Z926-Z928. Raw (noisy) data of the annular imploding shells prior to denoising is shown. The axes are in pixels. The conversion factor is 5 microns per pixel. Originally, before compression, the shells had millimeter size radii.

Ideally, one would acquire a sequence of such images per implosion using gated optics and track the evolution of a given accelerating shell.[4] Distortions in the shell as it implodes can be due to non-uniformities in the X ray illumination, target surface imperfections and their amplification due to hydrodynamic instabilities such as Rayleigh-Taylor.[1] From a given X ray backlighting image, one would like to extract the degree of asymmetry of the imploding shell near its peak plasma density or, equivalently, its radius of minimum X ray transmission.[3-10] Identifying that radius is the task at hand with appropriate shell thickness averaging and denoising. The main sources of the noise are thought to be the graininess of the film and the scanner digitization process.[3-10] This suggests that the noise is additive and the empirical evidence is that it is white Gaussian. For the very symmetric shot, Z927, and the intermediate one, Z926, this is certainly the case. For Z928, however, the noisiest and most distorted image analyzed herein (since it is taken later in time and therefore further along in the compression and thus has a much smaller radius) the log of the image was used and a Gaussian noise model seemed to fit that better. The techniques utilized in this paper should find fruitful application in the denoising of X ray backlit laser driven ICF target implosion images from facilities such as Omega[11] and the NIF.[12-14]

## 2. WAVELETS, CURVELETS AND DENOISING

In the past fifteen years, there has been a revolution in signal processing by the introduction and popularization of multiresolution analysis techniques and in particular, those based on wavelets.[15-18] The reign of Fourier as the dominant spectral representation domain where filtering, smoothing, scale identification and even phase space tiling concepts are evoked has been challenged by the advent of wavelets and their fast algorithmic implementation.[15-16] The successes of wavelets and their usage are too many to review here. A visit to www.wavelets.org will reveal the history of, tutorials and latest news concerning wavelets and their uses from astronomy to medicine, pure mathematics to audio engineering. For the purposes of this paper, the important tools are wavelets for image processing,[17-20] curvelets for image processing[21-23] and noise modeling (statistical analysis) and discrimination/thresholding techniques.[24-26]

The fundamental concept of multiresolution analysis is to decompose a signal or image into time (or space) and scale *simultaneously*. Thus, just as in music notation, not only which notes are to be played are specified (what Fourier supplies very precisely) but *when* and for what *durations* (about which Fourier is silent). The simultaneous scale and time (or space) decompositions are carried out in the simplest case by equispaced translates (whose number per scale or spacing is scale dependent) and $2^J$ dilates of a mother wavelet[15-16] (where J is the number of scales to be used). A wavelet is a localized waveform with zero mean and good localization properties both in space (time) and reciprocal space (frequency) (i.e. good localization in phase space). In signal processing terms, a wavelet is a band pass filter around a low pass filter used to detect the coarsest features of a signal (which itself is known as the scaling function in wavelet parlance). Wavelets record the finer and finer scale structures around the coarsest scale ones in a nested set of levels, with proportionately more translates at each finer scale. This pyramid structure can be constructed by the basic fast algorithm (O(N) operations, N being the number of data points, FFT being O(N lnN)) that was invented in the late eighties.[15-16] The resulting orthogonal[16] or biorthogonal[19] decompositions are said to be *decimated wavelet* ones. The drawback in denoising applications is that these decompositions are not translationally invariant while noise presumably is (how would the noise know which bit to corrupt?). The scale information content of a signal is the same no matter when we start or end the signal just as long as we keep all of it (all sequential permutations of the bits, that is). To adhere to that symmetry, the number of translates at each multiresolution level (scale) should not be changed and at all scales the number of translates should be the same as that at the finest scale. This gives rise to a highly redundant representation which nevertheless has great advantages in denoising since now the noise is diluted in a great many more coefficients only the most significant of which is to be kept via hard thresholding.[15-18] These constitute undecimated wavelet decompositions.[20] Denoising is done by estimating the variance of the noise and choosing to keep large wavelet coefficients which have a low probability of being noise. This is done iteratively as described in numerous references.[16-18, 24-26]

Turning to image processing, 1D wavelets can be used as tensor products. But this too has a major drawback. Wavelets are excellent edge detectors of a signal or point singularity or large derivative detectors. In 1D this is ideally suited for the task of identifying regions of significant change. In 2D, however, typical singularities or regions of significant change are not points but lines or curves. Edges in an image would be well localized by wavelets across the edge but not effectively along it. The highly correlated nature of points along a smooth curve does not require wavelets in order to be detected. Instead, extended objects with different orientations and scale dependent curvatures would be more suitable. It has been established that curves obeying parabolic scaling (i.e. ones whose widths scale as the square of their lengths) are ideally suited to detect edges in 2D images.[21-23] The implementation of these curvelet ideas has given rise to many recent successes in denoising images and will be used below to detect the minimum X ray transmission radii of noisy images. Curvelets will be seen to outperform both decimated and undecimated wavelet decompositions. The combined use of wavelets and curvelets with variational minimization to choose the best coefficients has also been seen to be beneficial in some cases.[23] Their application to imploding shell image denoising is of limited value and will be exposed elsewhere.[27-29]

This paper describes just one application of wavelets to fast Z pinch research and plasma physics. Other applications that we have already explored include denoising X ray bolometry and compressible turbulence data. In the former case, one is faced with the task of differentiating very noisy data, which is a Z pinch's X ray energy yield versus time, to obtain emitted X ray power vs time.[27-29] As for theoretical applications, two exciting areas are adaptive zoning in phase space for nonlinear kinetic theory (Vlasov) simulations,[30] and the compression of the significant degrees of freedom in describing radiation magnetohydrodynamic (RMHD) and compressible hydrodynamic (CHD) turbulence to a few wavelet coefficients.

## 2.1. 1D Wavelet Decompositions and (Naive) Denoising

The simplest multiresolution analysis based denoising to try with images such as those in Figure 2 is to center them as best as one can and to take radial cuts and denoise each cut separately using truncated wavelet series representations.[15-16] We show a few samples of the noisy and denoised radial cuts for Z926 and Z927 in Figures 3 and 4 including the histogram of minimum X ray transmission radii from 360 consecutive cuts in Figure 5.

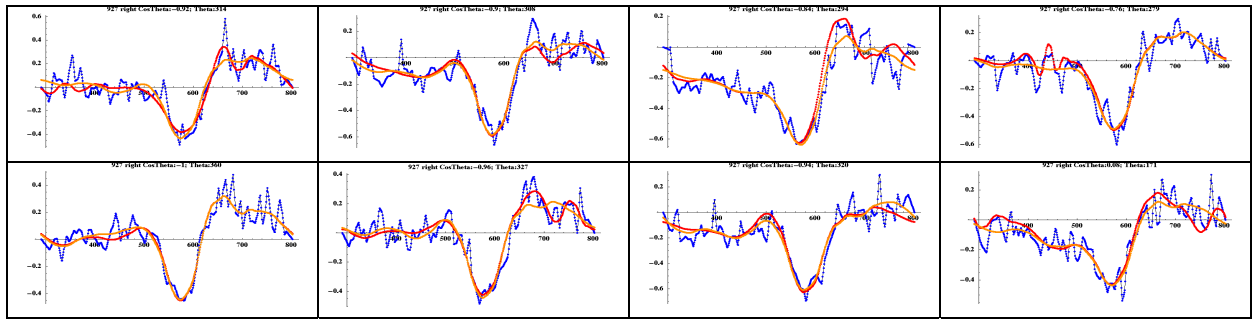

**Figure 3.** Radial cuts of the noisy Z927 image (blue) with lowest level (orange) and five largest coefficient (red) thresholding of the multiresolution decomposition using Daubechies 5 wavelets.

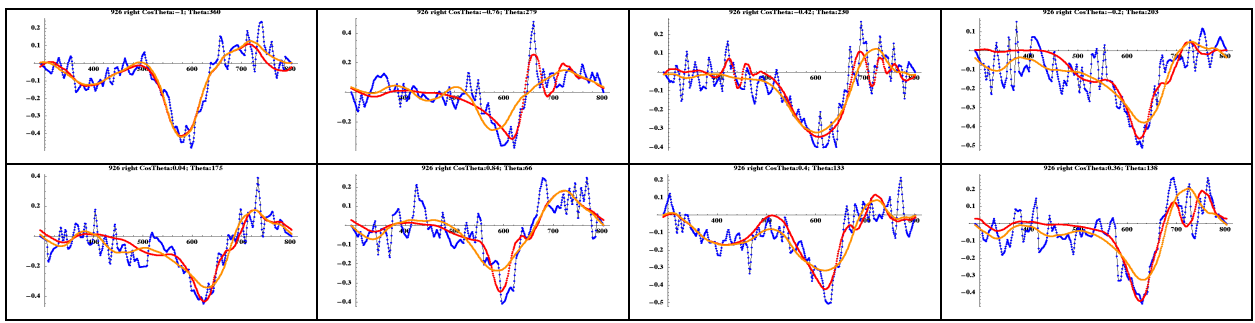

**Figure 4.** Radial cuts of the noisy Z926 image (blue) with lowest level (orange) and five largest coefficient (red) thresholding of the multiresolution decomposition using Daubechies 5 wavelets. Note that this is a much more asymmetric shell when compared to Z927 so that the lowest level and largest coefficient thresholding results no longer coincide. Nonlinear (largest coefficient, red) thresholding outperforms lowest level (orange) thresholding in tracking the minimum transmission radii.

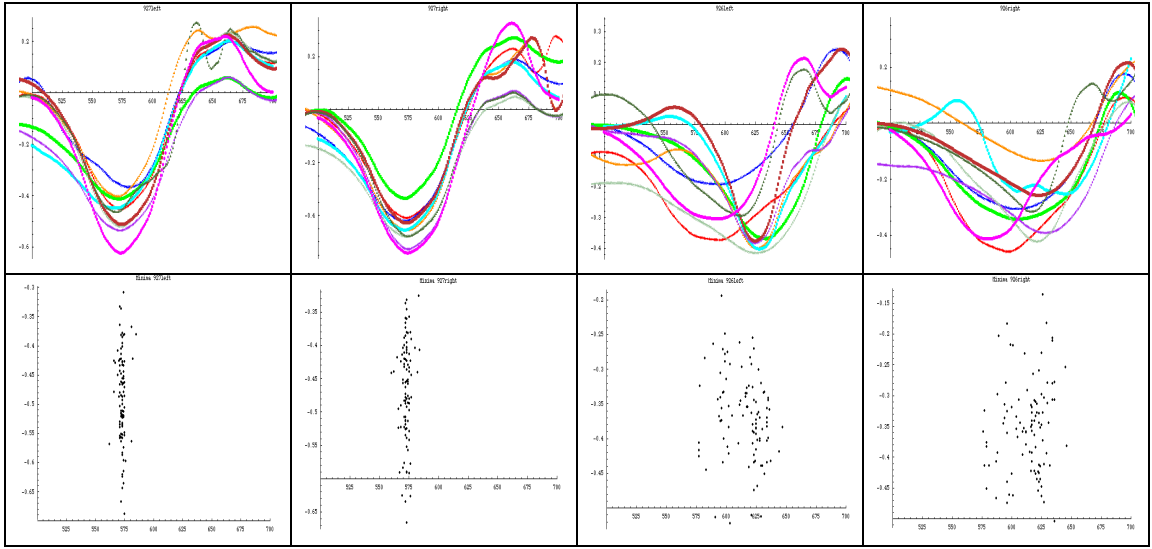

**Figure 5.** Denoised radial cut samples. Minimum X ray transmission radii for various angles, as inferred by 1D Daubechies 5 multiresolution decomposition and five largest wavelet coefficient thresholding. Radial cuts from the left and right semi-shells for different angles are shown separately. For Z927 (the first two columns) a smooth curve of $r_{min}(\theta)$ would result but not for Z926 (the last two columns). The bottom row consists of histograms of intensity minima vs their radial location for all angles sampled, some of which are shown in the figures 3 and 4.

The two denoised signals in Figs. 3- 4 correspond to linear thresholding by keeping the lowest level of the multiresolution decomposition using Daubechies 5 wavelets, and nonlinear thresholding by keeping the five largest Daubechies 5 coefficients only. In both cases, the results are very good, one radial cut at a time, and when the curve of minimum radius is not too distorted such as for Z927. But when it is strongly distorted, this method fails as the histograms for Z926 in figure 5 show. Whether the problem is dominated by proper centering, or actual pronounced shell distortions, or minimum transmission radii that are hard to pinpoint due to noise, it is difficult to distinguish by this signal-centric method. The method is naïve in another sense: it does not take into account the statistical properties of the noise in order to decide which wavelet coefficients to keep or reject. The third point of naivite stems from the fact that the minimum X ray transmission curve is strongly correlated in angle and individual radial cut denoising as shown in Figs. 3-5 is equivalent to assuming no angular correlation at all. Curvelets will address this last point. Denoising first and then centering combined with a Legendre Polynomial Decomposition (LPD) to characterize the angular nonuniformities of the imploding shell will be presented next using wavelets and curvelets.

## 2.2. 2D Wavelets, Curvelets Angular Correlation and (Noise Modeling Driven) Denoising

The images in figure 2 will be denoised using undecimated and decimated wavelets as well as curvelets. We have used the variance of the noise found in the individual images to estimate the coefficients which are likely to be noise dominated and discarding them which is referred to hard thresholding based on noise modeling.[18, 24-26] The decimated wavelet transform, being a non-redundant transform, does not allow for optimal denoising since it is not translationally invariant in its construction. Decimated biorthogonal wavelet transforms such as these are far more useful for image compression than for denoising.[19]

744     Proc. of SPIE Vol. 5207

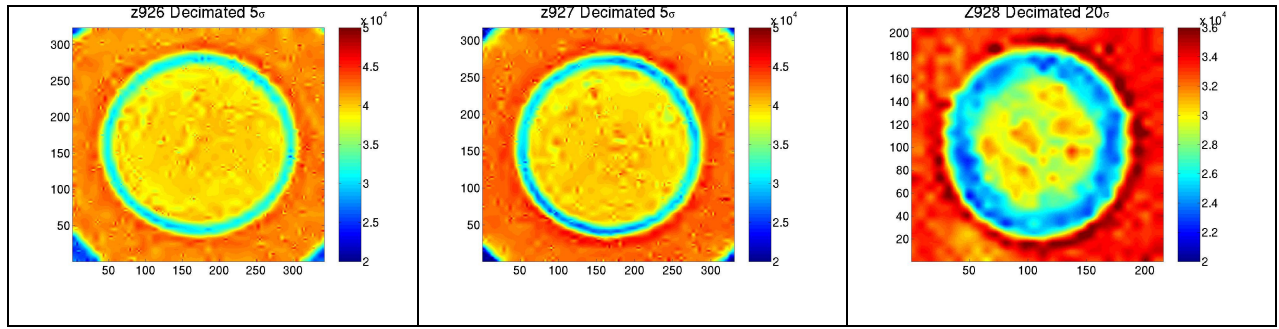

**Figure 6.** Decimated Wavelet transform reconstructions of X Ray backlighting image of double Z Pinch target shots Z926-Z928. The axes as depicted are in pixels. The conversion factor is 5 microns per pixel.

The next set of figures corresponds to a undecimated wavelet transform with hard thresholding based on noise modeling. This highly redundant transform does allow for optimal denoising since it is translationally invariant in its construction. Discrimination against noise is much easier in this transform than in the decimated transform case. However, many artifacts still remain since a point wise and isotropic construction is being implemented when wavelets are used. To adapt to the contours of the figure a truly 2D construction is needed such as that afforded by curvelets.[21-23] Curvelets afford a very useful tool for denoising images such as these. Figure 8 shows this directly.

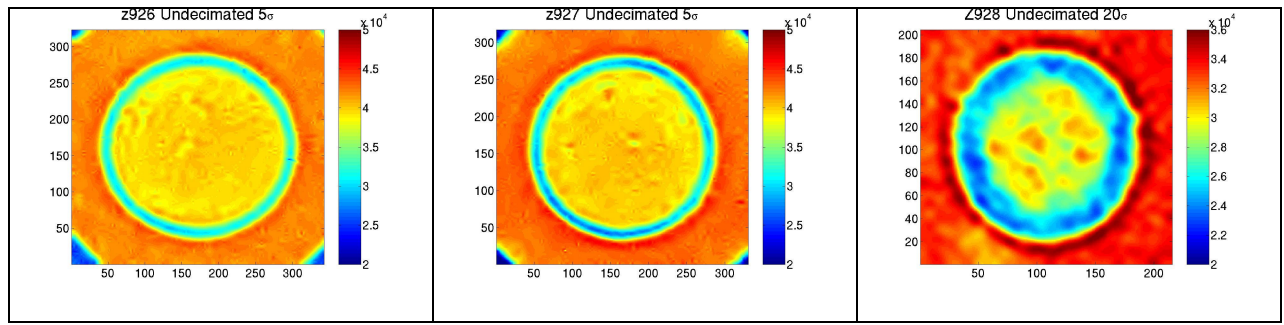

**Figure 7.** Undecimated Wavelet transform reconstructions of X Ray backlighting image of double Z Pinch target shots Z926-Z928. The axes as depicted are in pixels. The conversion factor is 5 microns per pixel.

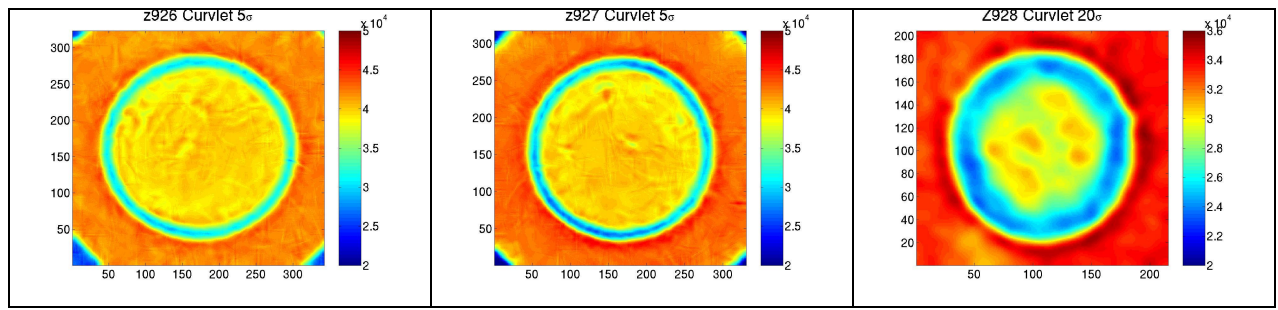

**Figure 8.** Curvelet transform reconstructions of X Ray backlighting image of double Z Pinch target shots Z926-Z928. The clarity of the annular shell is apparent in these images. The axes as depicted are in pixels. The conversion factor is 5 microns per pixel.

One way to see just how the imploding shells behave as a function of angle is to plot the locus of the radius of *minimum* X ray transmission sandwiched between the loci of the inner and outer radii of *maximum* transmission. These max-min-max polar plots are shown in figures 9-11 for the denoised and reconstructed images using the three filters, decimated and undecimated wavelets as well as curvelets. The asymmetry of the implosion can be easily seen from the nonuniformity of the distance separating the various curves. Z927 is the most symmetric while Z928 is the least. The wavelet decompositions (decimated or undecimated) do not allow a clear demarcation of the min and max radii for the highly asymmetric Z928 case. That is why we include it in the curvelet decomposition case only where the loci are identified successfully.

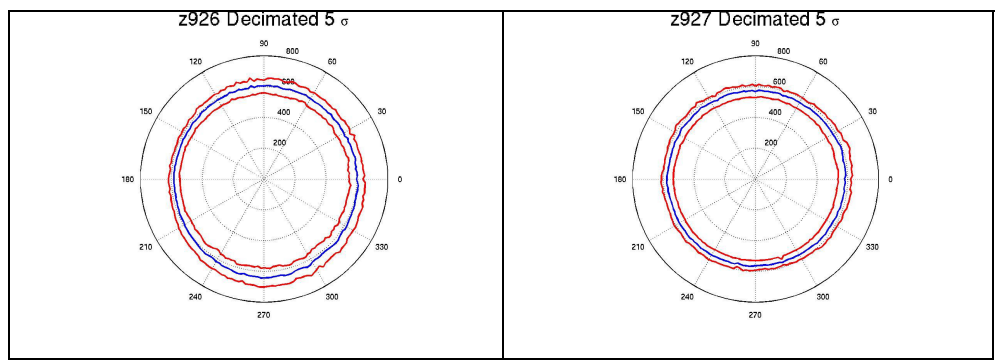

**Figure 9.** Polar max-min-max plots of minimum and maximum X ray transmission radii for decimated wavelet transform reconstructed X ray backlighting images of Z926 and Z927. Z928 is not shown because it would have a large number of exaggerated radial spikes missing a smooth min curve entirely.

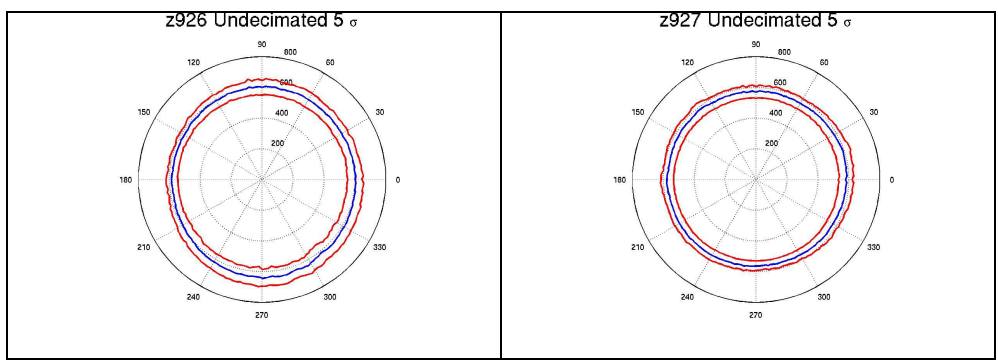

**Figure 10.** Polar max-min-max plots of minimum and maximum X ray transmission radii for undecimated wavelet transform reconstructed X ray backlighting images of Z926-Z927. Z928 is not shown because it would have a large number of pronounced radial spikes missing a smooth min curve entirely.

The curvelet decompositions produce the smoothest min and max loci as seen in Fig. 11. We can now take our first glimpse as to how distorted Z928 really is. Further analysis will quantify this below.

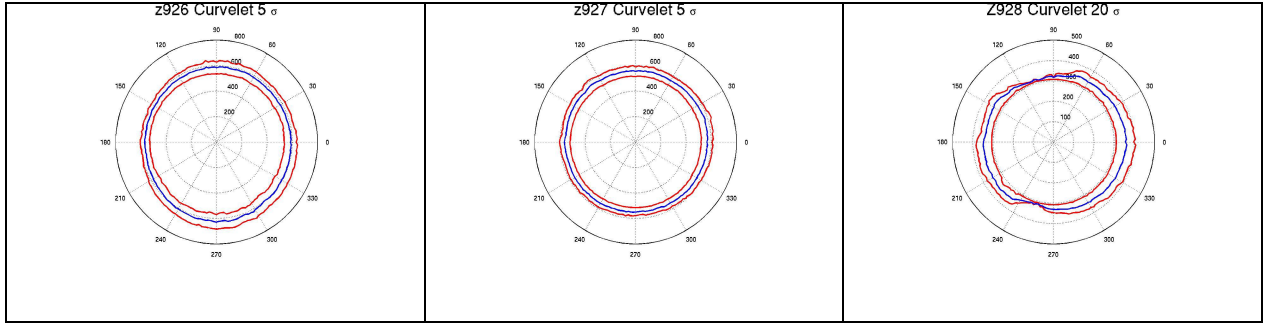

**Figure 11.** Polar max-min-max plots of minimum and maximum X ray transmission radii for curvelet transform reconstructed X ray backlighting images of Z926-Z928. The reason the Z928 min and max curves seem to pinch around a 100 degrees and again around 250 is because the shells seem to be flattened there making the concept of min and max blurred.

A better way to characterize the deviations of a shell from polar symmetry (in its projection onto a plane), is to unwrap the circular ribbon like structures by plotting the images as radius vs angle on a Cartesian grid. Now a shell becomes a ribbon, a circle becomes a horizontal line and any deviations from that line imply distortions and fluctuations. Nested in the middle of the ribbons, we also plot r*(0, θ), r*(50, θ) and r*(90, θ) where the first argument of r* is the percentage of the distance between the minimum transmission radius and its closest maximum (on either side of the minimum). This is the shell thickness averaged radius (STAR), r*, where I(r, θ) is the X ray intensity distribution in the image:

$$r^*(N,\theta) = \int_{r_L(N,\theta)}^{r_R(N,\theta)} I(r,\theta) r\, dr \bigg/ \int_{r_L(N,\theta)}^{r_R(N,\theta)} I(r,\theta)\, dr$$

$$I(r_L) = I(r_R) = \frac{N}{100}\left(Min[I(r_{max,L}), I(r_{max,R})] - I(r_{min})\right)$$

We have found that r*(90, θ) has the smoothest and most useful qualities of the three shown in Figures 12-14.

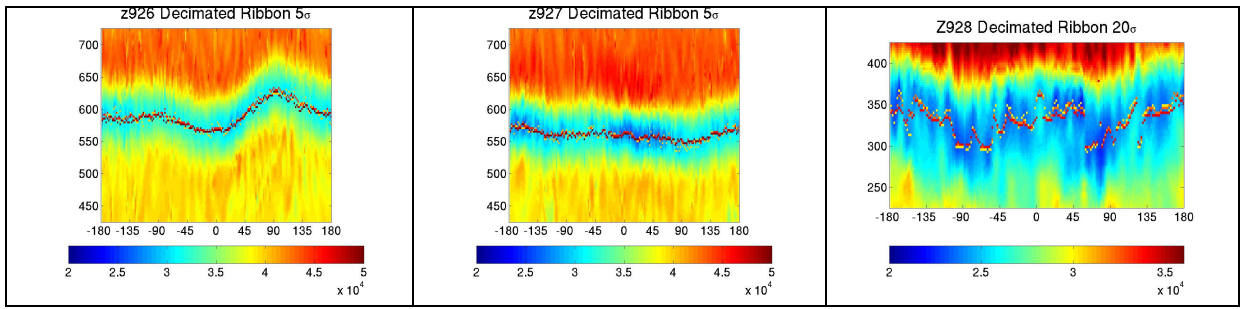

**Figure 12.** Interpolated polar coordinates ribbon plots: Decimated Wavelet transform reconstructions of X Ray backlighting image of double Z Pinch target shots Z926-Z928 with r*(0), r*(50) and r*(90) superposed.

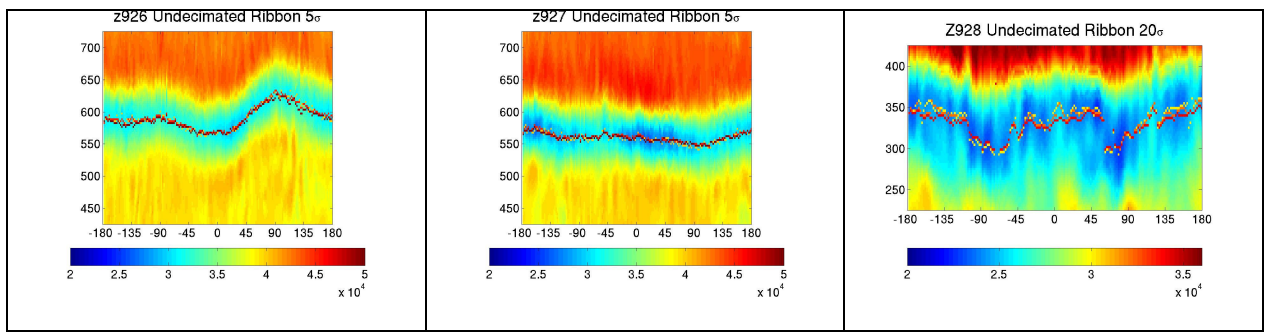

**Figure 13.** Interpolated polar coordinates ribbon plots: Undecimated Wavelet transform reconstructions of X Ray backlighting image of double Z Pinch target shots Z926-Z928 in unwrapped ribbon format with r*(0, θ), r*(50, θ) and r*(90, θ) superposed.

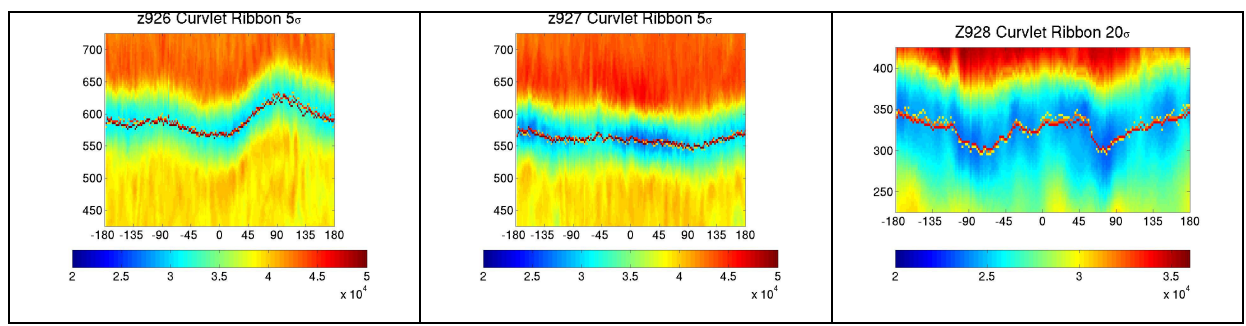

**Figure 14.** Interpolated polar coordinates ribbon plots: Curvelet transform reconstructions of X Ray backlighting image of double Z Pinch target shots Z926 with r*(0, θ), r*(50, θ) and r*(90, θ) superposed.

Next, in Figures 15-16, we plot the r*(90, θ) curves (without the ribbons), obtained by using decimated and undecimated wavelet and curvelet filters. For Z928, only curvelets produce a smooth r*(90, θ) curve. The wild variations between the various cases with different filters for Z928 shows just how important it is to use a sensible filter and not to succumb to the commonly held notion that any one of them will do since they give more or less the same result. Fig. 16 shows the ratio of r* to its average. We will Legendre decompose r*(90, θ) obtained via curvelets next.

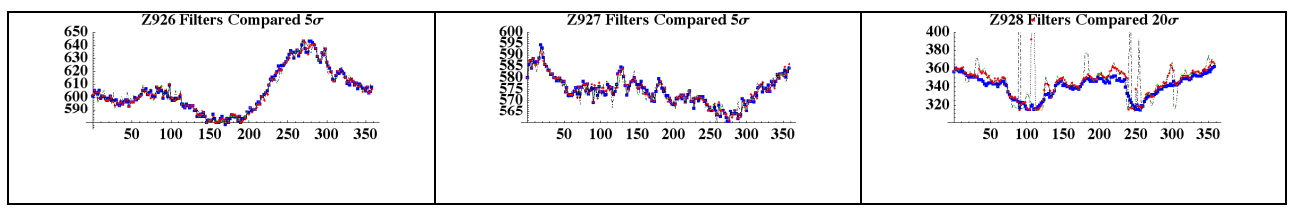

**Figure 15.** A comparison of the r*(90, θ) curves obtained via the three filters on the three data sets. Black (dashed) is via a decimated wavelet transform, red (diamonds), undecimated, and blue (squares) corresponds to curvelets. The wavelet transformed ones do no function properly for Z928 since they cannot track the minimum causing spiking or walk off. This is anticipated from the polar min-max plots shown above (for curvelets only) where for the same angles (around 100 and 250, the min and max curves almost intersect not allowing for a clean min or averaged min extraction). In all cases, curvelets produce the least noisy contours.

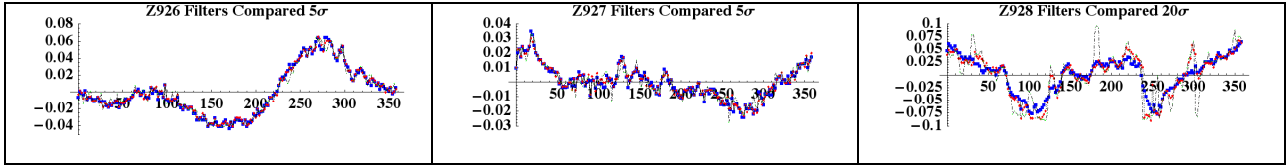

**Figure 16.** A comparison of the r*(90, θ) curves obtained via the three filters on the three data sets each normalized to its own average radius. The average radii for the three Z images are roughly 603, 574 and 340, respectively. Black (solid lines) corresponds to a decimated wavelet decomposition, red (diamonds), undecimated, and blue (squares) corresponds to a curvelets based decomposition. Note that only curvelets can detect the R*(90, θ) curve for Z928. Wavelets fly off the mark.

Finally, the Legendre polynomial decompositions of curvelet generated r*(90, θ) curves for Z926-Z928 are given in Figure 17. The degrees of asymmetry in the imploding shells are thus quantified. The vertical axes contain the Legendre polynomial coefficients $c_n$ normalized to the zero order Legendre polynomial coefficient $c_0$. $c_0$ is equal to the average diameter of the curve r*(90, θ). The coefficients $c_n$ are found by Legendre decomposing each half of r* separately, since the integrals are over cos θ in the interval [-1,1]. The two semi-circular decomposition results are then added.

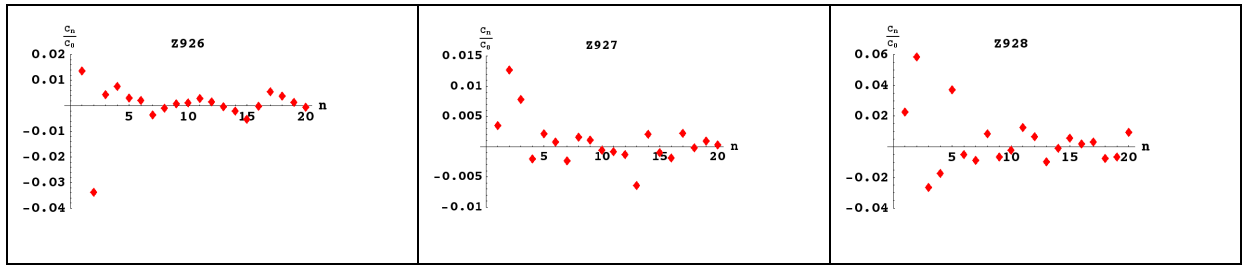

**Figure 17.** The Legendre polynomial decomposition of curvelet generated r*(90) curves for Z926-Z928. The first 20 mode amplitudes are shown. We see that Z927 has around 1.5% $P_2$ and 0.2% $P_4$ deformations, Z926 has over 3.5% $P_2$ and almost 1% $P_4$ deformations while Z928 has over 6% $P_2$ and 2% $P_4$ deformations. These are decompositions without any attempts at centering the images to better than plus or minus 5 pixels of the true centers. That is why the $P_1$ contributions are nonzero but small in all three cases.

We have compared the LPD results from r*(90, θ) curves produced by wavelets, curvelets and total variation diminishing and $L_1$ norm diminishing combined undecimated wavelet and curvelet representations[23] in two other publications.[28-29] Better centering, image registration, active contours, snakes and contourlets are just some of the avenues of attack yet to be explored. The exciting prospect exists that imploding shells in ICF can be diagnosed quantitatively by X ray backlighting so that small deformations and asymmetries can be unambiguously detected once the images are properly denoised.

## ACKNOWLEDGMENTS

This work was supported by a contract with Sandia National Laboratories. The authors would like to thank David Donoho for his constructive criticism and Ken Struve, Dillon McDaniel, John Porter, Jim Hammer, Roger Vesey, Guy Bennett and Dan Sinars for their valuable input and encouragement.

## REFERENCES

1. J. H. Hammer, et al., "High yield inertial confinement fusion target design for a Z-pinch-driven hohlraum," *Phys. Plasmas* **6**, 2129, 1999.
2. R. E. Olson, et al., "Indirect-drive ICF target concepts for the X-1 Z pinch facility," *Fusion Technology* **35**, 260, 1999.
3. M. E. Cuneo, et al., "Development and characterization of a Z-pinch-driven hohlraum high-yield inertial confinement fusion target concept," *Phys. Plasmas* **8**, 2257, 2001.
4. G. R. Bennett, et al., "X-Ray imaging techniques on Z using the Z-Beamlet laser," *Rev. Sci. Inst.* **72**, 657, 2001.
5. D. L. Hansen, et al., "Measurement of radiation symmetry in Z-pinch-driven hohlraums," *Phys. Plasmas* **9**, 2173, 2002.
6. M. E. Cuneo, et al., "Double Z-pinch hohlraum drive with excellent temperature balance for symmetric inertial confinement fusion capsule implosions," *Phys. Rev. Lett.* **88**, 215004, 2002.
7. G. R. Bennett, et al., "Symmetric inertial confinement fusion capsule implosions in a double Z pinch driven hohlraum," *Phys. Rev. Lett.* **89**, 245002, 2002.
8. R. A. Vesey, et al., "Demonstration of radiation symmetry control for inertial confinement fusion in double Z-pinch hohlraums," *Phys. Rev. Lett.* **90**, 035005, 2003.
9. G. R. Bennett, et al., "Symmetric inertial confinement fusion capsule implosions in a high yield scale double Z pinch driven hohlraum on Z," *Phys. Plasma* **10**, 3717, 2003.
10. R. A. Vesey, et al., "Radiation symmetry control for inertial confinement fusion capsule implosions in double Z-pinch hohlraums on Z," *Phys. Plasmas* **10**, 1854, 2003.
11. P. A. Amendt, R. E. Turner and O. L. Lander, "Hohlraum-driven high convergence implosion experiments with multiple beam cones on the Omega laser facility," *Phys. Rev. Lett.* **89**, 165001, 2002.
12. O. L. Landen et al., "X ray backlighting for the National Ignition Facility," *Rev. Sci. Inst.* **72**, 627, 2001.
13. S. M. Pollaine, et al., "National Ignition Facility scale hohlraum asymmetry studies by thin shell radiography," *Phys. Plasmas* **8**, 2357, 2001.
14. P. A. Amendt et al., "Indirect drive non-cryogenic double-shell ignition targets for the national ignition facility: Design and analysis," *Phys. Plasmas* **9**, 2221, 2002.
15. I. Daubechies, *Ten Lectures on Wavelets*, SIAM Press, Philadelphia, PA, 1992.
16. S. Mallat, *A Wavelet Tour of Signal Processing*, 2$^{nd}$ Edition, Academic Press, San Diego, CA, 1999.
17. J. L. Starck, F. Murtagh and A. Bijaoui, *Image Processing and Data Analysis: the Multiscale Approach,* Cambridge University, Cambridge, UK, 1998.
18. J. L. Starck and F. Murtagh, *Astronomical Image and Data Analysis,* Springer Verlag, Berlin, 2002.
19. M. Antonini, et al., "Image coding using wavelet transforms," *IEEE Trans. Image Processing*, **1**(2), 205, 1992.
20. M. Shensa, "The discrete wavelet transform: wedding the a trous and Mallat algorithms," *IEEE Trans. Signal Processing*, **40**(10), 2464, 1992.
21. J. L. Starck, et al., "The curvelet transform for image denoising," *IEEE Trans. Image Processing*, **11**(6), 670, 2002.
22. J. L. Starck, et al., "Astronomical image representation by the curvelet transform," Astronomy & Astrophysics **398**, 785, 2003.
23. J. L. Starck, D. L. Donoho and E. J. Candes, "Very high quality image restoration by combining wavelets and curvelets," Proceedings SPIE **4478**, 9, 2001.
24. A. Antoniadis and G. Oppenheim, eds., *Wavelets and Statistics,* Springer-Verlag, New York, 1995.
25. W. Hardle, G. Kerkyacharian, D. Picard and A. Tsybakov, eds., *Wavelets, Approximation and Statistical Applications*, Springer-Verlag, 1998.
26. M. Jansen, *Noise Reduction by Wavelet Thresholding*, Springer-Verlag, 2001.
27. B. Afeyan, K. Won, J. L. Starck, M. E. Cuneo, G. R. Bennett and R. A. Vesey, "Applications of wavelets to signal and image processing in fast Z pinch research," to be submitted to *Rev. Sci. Inst.*, 2003.
28. B. Afeyan, K. Won, J. L. Starck and M. E. Cuneo, "Wavelets, curvelets and multiresolution analysis techniques applied to implosion symmetry characterization of ICF targets," Proceedings of IFSA, Sept. 7-12, Monterey, CA, 2003.
29. K. Won, et al., "The true symmetry of double Z pinch driven imploding shells using multiresolution denoising techniques on X ray backlighting generated images," Bulletin of the American Physical Society, Division of Plasma Physics, Annual Meeting, Albuquerque, NM, Oct. 27-31, 2003, and to be published.
30. B. Afeyan, K. Won, V. Savchenko and J. L. Starck, "Wavelet and multiresolution techniques applied to adaptive, non-uniform phase space tiling for Vlasov simulations," to be submitted to *J. Comp. Phys.*, 2003.